\documentstyle[12pt,eqsecnum,aps,prc,floats,graphicx,tighten]{revtex}

\def\nuc#1#2{${}^{#1}$#2}

\begin{document}

\title{Track Restore Technique (RST) Applied to Analysis of Waveform
\\of Voltage Pulse in SAGE Proportional Counters}

\author{J. N. Abdurashitov, T. V. Ibragimova, A. V. Kalikhov}
\address{\it Institute for Nuclear Research, Russian Academy of Sciences,
117312 Moscow, Russia}
\author{(for the SAGE collaboration)}
\maketitle

\begin{abstract}
A kind of analysis of waveform of voltage pulse in small proportional
counter is described. The technique is based on deconvolution of recorded
pulse with a response function of the counter. It allows one to restore
a projection of the track of initial ionization to the radius of the counter,
or a charge collection function.
\end{abstract}

\section{Introduction}

One of main procedures of solar neutrino flux measurement in gallium experiments
is a detection of several atoms of \nuc{71}{Ge} in a small proportional counter.
\nuc{71}{Ge} decays solely via electron capture to the ground
state of \nuc{71}{Ga}. In the proportional counter 1.2 keV and 10.4 keV
Auge-electrons are usually detected. These low-energy electrons produce
a nearly point-like ionization in the counter gas. This ionization will
arrive at the anode wire of the proportional counter as a unit resulting
in a fast rise time for the pulse. In contrast, although a typical
$\beta$-particle produced by a background process may also lose 1 keV to
15 keV in the counter gas, it will leave an extended trail of ionization.
The ionization will arrive at the anode wire distributed in time according
to its radial extent in the counter, which usually gives a pulse with a slower
rise time than for a \nuc{71}{Ge} event. The identification of true \nuc{71}{Ge} events
and the rejection of background events is thus greatly facilitated by using
a two parameter analysis: a candidate \nuc{71}{Ge} event must not only fall
within the appropriate energy region, but must also have a rise time consistent
with point-like ionization.

\begin{figure}
\begin{center}
\includegraphics[width=4in]{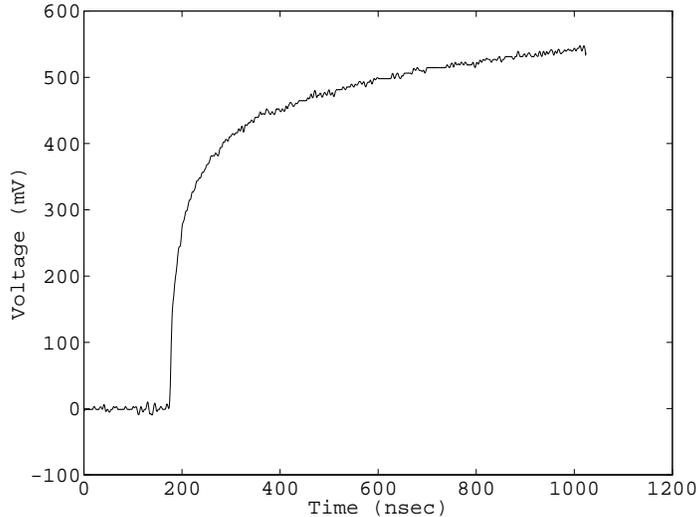}
\end{center}
\caption{An example of typical point-like pulse produced by 10.4-keV Auge-electron
after \nuc{71}{Ge} decay.}
\label{Kpoint}
\end{figure}

The anode wire is directly connected to a charge-sensitive preamplifier. After
further amplification the signal is going (in SAGE) to the digital oscilloscope
HP5411D, which records the voltage pulse waveform with 8-bit voltage resolution
and 1 ns time resolution for 800 ns after pulse onset. A typical pulse produced
by 10.4 keV Auge-electron after \nuc{71}{Ge} decay is shown on Fig.\ \ref{Kpoint}.

\section{Standard analysis of waveform}

There are several different techniques which are applied to waveform analysis
in both gallium solar neutrino experiments. All of them are described in detail
elsewhere (see ~\cite{sre}, ~\cite{lng} for SAGE,  ~\cite{alt} for GALLEX).
For example, in the standard analysis of SAGE data so called $T_N$ method
is used there. A functional form described in ~\cite{sre} with parameter $T_N$
characterizing rise time of the pulse is written for fit the
observed pulse to.
This technique gives the correct
description of the shape of the voltage pulse as recorded by the digital
oscilloscope when the ionization produced in the proportional counter
consists of a set of point ionizations evenly distributed along a straight
track.  Since \nuc{71}{Ge} events are usually a single cluster of ionization,
this method works satisfactorily to select \nuc{71}{Ge} candidate events.  It
is, however, restricted to the particular form of ionization that is assumed,
and gives a poor fit to other types of charge deposit in the counter, such as
the combination of a point event from \nuc{71}{Ge} $K$-electron capture
followed by capture of the 9.3-keV x ray at some other location in the
counter.  To give us the capability to investigate all possible events that
may occur in the counter, we have also developed a more general method which
can analyze an event produced by ionization with an arbitrary distribution of
charge.  We call this the `restored pulse method', or `RST method' for short.

\section{Description of RST technique}

We begin with the measured voltage pulse $V(t)$ as recorded by the
digitizer.  For an ideal point charge that arrives at the counter anode wire,
$V(t)$ has the Wilkinson form $V(t) = W(t) = V_0 \ln(1 + t/t_0)$, provided
the counter is ideal and the pulse processing electronics has infinite
bandwidth.  For a real event from the counter, with unknown charge
distribution, $V(t)$ can in general be expressed as the convolution of the
Wilkinson function with a charge collection function $G(t)$:
\begin{equation}
\label{G_definition}
V(t) = W(t) \otimes G(t).
\end{equation}
The function $G(t)$ contains within it the desired information about the
arrival of charge at the counter anode, coupled with any deviations of the
counter or electronics from ideal response.  Equation~(\ref{G_definition})
can be considered as the definition of $G(t)$.

To get the desired function $G(t)$, one must deconvolute
Eq.\ (\ref{G_definition}).   To perform this deconvolution, we have found it
mathematically convenient to use the current pulse $I(t)$, which is obtained
by numerical differentiation of $V(t)$:
\begin{eqnarray}
I(t) & = & \frac{dV}{dt} = \frac{d}{dt} (W(t) \otimes G(t)) \\ \nonumber
     & = & \frac{dW}{dt} \otimes G(t) = W^{'}(t) \otimes G(t),
\end{eqnarray}
where $W^{'}(t)$ is normalized over the observed time of pulse measurement
$T_{\text{obs}}$ such that $\int_0^{T_{\text{obs}}} W^{'}(t)dt = 1$.

To deconvolute, we Fourier transform to the frequency domain and then
use the theorem that convolution in the time domain becomes multiplication in
the frequency domain.  This simply gives $I(f) = W^{'}(f)
G(f)$, which can be solved for $G(f)$.  We then Fourier transform $G(f)$ back
to the time domain to get the desired function $G(t)$.  The energy of the
event is given by $\int_0^{T_{\text{obs}}} G(t) dt$.  The duration of the
collection of ionization is given by the width of $G(t)$, which can be used
as a measure of the rise time.

\begin{figure}
\begin{center}
\includegraphics[width=4in]{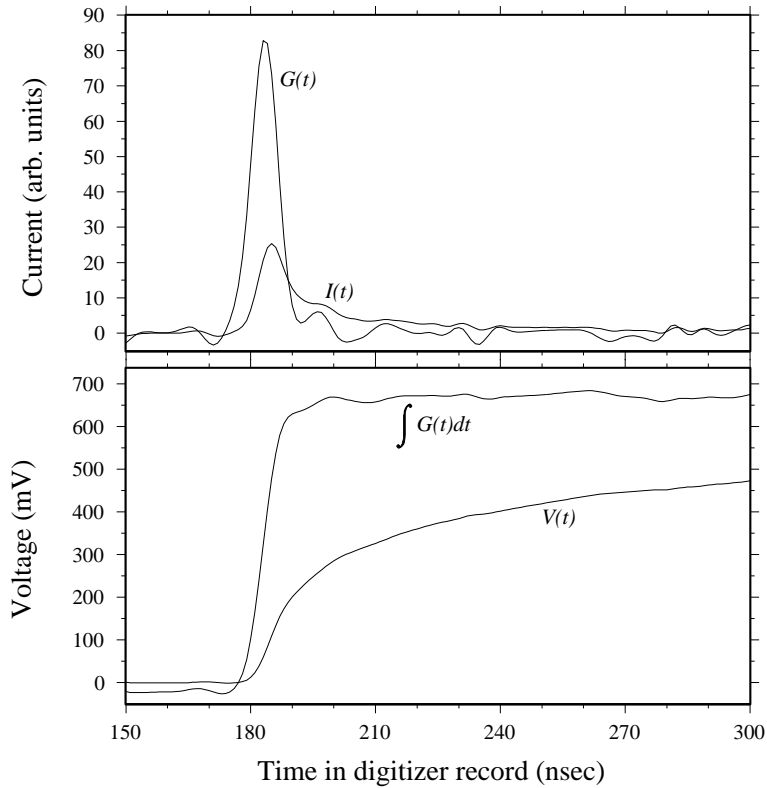}
\end{center}
\caption{Analysis of typical \nuc{71}{Ge} pulse by the RST method.  See text
for explanation.}
\label{RST_example}
\end{figure}

An example of this procedure as applied to a typical \nuc{71}{Ge}
$K$-peak event is given in Fig.\ \ref{RST_example}.  This pulse has $T_N =
3.9$ ns.  The recorded voltage pulse after inversion and smoothing is given
by $V(T)$ in the lower panel.  The current pulse, obtained by numerical
differentiation of the voltage pulse, is given by $I(t)$ in the upper panel.
The deduced function $G(t)$ is also shown in the upper panel.  It has a FWHM
of about 15 ns, found to be typical for true \nuc{71}{Ge} $K$-peak events.
The integrated current pulse, which records the pulse energy, is given by
$\int G(t)dt$ in the lower panel.

\section{Conclusion}

This method has the advantage that it can reveal the basic nature of the
ionization in the counter for an arbitrary pulse.  It is also capable of
determining the pulse energy over a wider range than the $T_N$ method.  A
problem that has been found with this method in practice, however, is that
when \nuc{71}{Ge} data are analyzed one obtains multiple collection functions
(i.e., $G(t)$ has several distinct peaks separated in time) more often than
is expected from the known physical processes that take place in the counter.
These multiple peaks are due to noise on the pulse and cutoff of the system
frequency response at about 100 MHz.  Attempts have been made to remove these
extraneous peaks by filtering and smoothing the original pulse, but they have
not been fully successful.  Evidently we need faster electronics and a
reduction in the noise level to be able to fully exploit this pulse shape
analysis technique.  As a result, we have only been able to use this method
to select events on the basis of energy.

\section{Acknowledgments}

We thank many members of SAGE for fruitful and stimulating discussions.
Especially we thank B. T. Cleveland for his help in careful preparation
of the article.


\begin{thebibliography}{3}

\bibitem{sre} S. R. Elliott,
              Nucl.\ Instrum.\ Meth.\ in Phys.\ Res.\ A {\bf 290}, 158 (1990).
\bibitem{lng} J. N. Abdurashitov, V. N. Gavrin, S. V. Girin {\em et al.},
              astro-ph/9907113
\bibitem{alt} M. Altmann, F. v. Feilitzch, U. Schanda,
              Nucl.\ Instrum.\ Meth.\ in Phys.\ Res.\ A {\bf 381}, 398 (1996).

\end{thebibliography}
\end{document}